\documentclass[aps,pre,twocolumn,10pt,superscriptaddress,nofootinbib,balancelastpage]{revtex4-1} 
\usepackage[latin1]{inputenc}
\usepackage{amsmath,amssymb}
\usepackage{mathrsfs} 
\usepackage[capitalise]{cleveref}
\usepackage{siunitx}
\usepackage{braket}
\usepackage{calc}
\usepackage{tabularx}
\usepackage{dsfont}
\usepackage{color}
\usepackage{ifthen}
\usepackage{graphicx}
\usepackage{seqsplit}
\allowdisplaybreaks

\newcommand{\Eins}{\mathds{1}}%

\newcommand{\ii}{\mathrm{i}}%
\newcommand{\dd}{^{(3)}}
\newcommand{\organized}{\mathfrak{v}}

\newcommand{\dif}{\mathrm{d}}%
\newcommand{\pdif}[2]{\frac{\partial#1}{\partial#2}}%
\newcommand{\Nabla}{\vec{\nabla}}%
\newcommand{\fdif}{\operatorname{\delta}}%
\newcommand{\Fdif}[2]{\frac{\fdif\!#1}{\fdif\!#2}}%

\newcommand{\vorf}{\frac{1}{\rho_{\mathrm{l}}}}

\newcommand{\rt}{(\vec{r},t)}

\newcommand{\Tr}{\operatorname{Tr}}%
\newcommand{\ZT}[1]{\textquotedblleft#1\textquotedblright}%

\newcolumntype{Y}{>{\centering\arraybackslash}X}%     
\newcolumntype{Z}{>{\raggedright\arraybackslash}X}%   

\newlength{\myl}%
\newcommand{\SUM}[2]{{\setlength{\myl}{\widthof{$\displaystyle\sum_{#1}^{#2}$}*\real{0.5}-\widthof{$\displaystyle\sum$}*\real{0.5}}\sum_{#1}^{#2}\;\hspace{-\the\myl}}}% Summen in abgesetzten Gleichungen
\newcommand{\INT}[3]{\settowidth{\myl}{$\displaystyle\int_{#1}^{#2}$}{\int_{#1}^{#2}\;\;\;\hspace{-\the\myl}\dif #3}\,}% Integrale in abgesetzten Gleichungen
\newcommand{\TINT}[3]{\settowidth{\myl}{$\int_{#1}^{#2}$}{\int_{#1}^{#2}\!\ifthenelse{\equal{#1#2}{}}{}{\;\;\;\;\hspace{-\the\myl}}\dif #3}\,}%
\newcommand{\EINT}[3]{\settowidth{\myl}{$\int_{#1}^{#2}$}{\int_{#1}^{#2}\;\;\;\,\hspace{-\the\myl}\dif #3}\,}% Integrale in Exponenten

\newcommand{\eps}[1]{\epsilon_{\text{#1}}}

\newcommand{\randomf}{\mathfrak{z}}

\DeclareMathOperator\erf{erf}

\begin{document}
\title{Microscopic derivation of the thin film equation using the Mori-Zwanzig formalism}

\author{Michael te Vrugt}
\affiliation{DAMTP, Centre for Mathematical Sciences, University of Cambridge, Cambridge CB3 0WA, United Kingdom}

\author{Leon Topp}
\affiliation{Institute of Physical Chemistry, Universit\"at M\"unster, 48149 M\"unster, Germany}

\author{Raphael Wittkowski}
\email[Corresponding author: ]{raphael.wittkowski@uni-muenster.de}
\affiliation{Institute of Theoretical Physics, Center for Soft Nanoscience, Universit\"at M\"unster, 48149 M\"unster, Germany}

\author{Andreas Heuer}
\affiliation{Institute of Physical Chemistry, Universit\"at M\"unster, 48149 M\"unster, Germany}

\begin{abstract}
The hydrodynamics of thin films is typically described using phenomenological models whose connection to the microscopic particle dynamics is a subject of ongoing research. Existing methods based on density functional theory provide a good description of static thin films, but are not sufficient for understanding nonequilibrium dynamics. In this work, we present a microscopic derivation of the thin film equation using the Mori-Zwanzig projection operator formalism. This method allows to directly obtain the correct gradient dynamics structure along with microscopic expressions for the mobility and the free energy. Our results are verified against molecular dynamics simulations for both simple fluids and polymers.
\end{abstract}
\maketitle

\section{Introduction}
In recent years, a significant amount of work has been done on the description of thin films on substrates. This includes a variety of aspects, ranging from contact line motion \cite{FrickeMB2020,FrickeMB2019} and pattern formation \cite{TewesWGT2019,HonischLHTG2015} to sliding droplets \cite{LiEtAl2022,WongBLVSWB2022,WilczekTEGT2017,EngelnkemperWGT2016}. An improved understanding of thin films is of interest, e.g., for technological applications  \cite{HanZ2012,WeinsteinR2004,LyMTG2020,LyTCG2019,MitasMT2020,KasischkeHNKTG2021} or the modeling of bacterial colonies \cite{TrinschekSJT2020,TrinschekJLT2017,TrinschekJT2018}. An improved understanding of thin films is therefore of interest in a large variety of disciplines, including biology \cite{WallmeyerTYTB2018}, chemistry \cite{BoeckmannSdJDHD2015}, engineering \cite{ParkKXW2018}, mathematics \cite{FrickeKB2018}, and physics \cite{CrasterM2009,BonnEIMR2009}. Many applications require an understanding not only of the equilibrium configuration of thin films but also of their dynamics out of equilibrium.

Two main modeling approaches can be distinguished. First, a microscopic description is possible using particle-based simulations \cite{LeonforteSPM2011,MilchevB2001,DongSS1996}. Second, thin films can be described on a macroscopic level using continuum models, which often have a variational form \cite{Mitlin1993,Thiele2018,WilczekTGKCT2015,Thiele2014,Thiele2012,ThieleH2020}. The latter can be derived either as an approximation to the Navier-Stokes equation via a long-wave approximation \cite{Engelnkemper2017} or from gradient dynamics based on thermodynamic arguments \cite{ThieleH2020}. An understanding of the connection between microscopic and macroscopic approaches is of high interest, both for the interpretation of these models and for the development of extensions.

Various approaches have been developed to establish a connection between particle-based and continuum approaches \cite{TewesBHTG2017,BullerTAHTG2017,TretyakovMTT2013,StienekerTGH2021,ToppSGH2022}. A very useful framework in this context is density functional theory (DFT) \cite{YinSA2019,YatsyshinDK2018,HughesTA2017,YatsyshinK2016,YatsyshinK2016b,YatsyshinSK2015,HughesTA2014,HughesTA2015,LiW2008,YatsyshinSK2015b,NoldSGK2014}, which allows to find the equilibrium configuration of a fluid based on minimizing a free energy functional. The nonequilibrium dynamics of thin films can be modeled in dynamical density functional theory (DDFT) \cite{RobbinsAT2011,ThieleVARFSPMBM2009,ArcherRT2010,ChalmersSA2017,HowardNP2017,HowardNP2017b,SearW2017,ThieleAP2012,AlandV2012,GrawitterS2018,YeTZDM2016,HuininkBvDS2000,HorvatLSZM2004,TsarkovaHKZSM2006,MoritaKD2001,ParadisoFFF2012}, which is the nonequilibrium extension of DFT \cite{Evans1979,MarconiT1999,ArcherE2004} (see Ref.\ \cite{teVrugtLW2020} for a recent review). However, DFT is restricted to the equilibrium case and DDFT to diffusive dynamics, such that both approaches do not provide a full picture of the nonequilibrium dynamics of thin films. 

A systematic connection between microscopic and macroscopic descriptions of physical systems can be obtained using the Mori-Zwanzig projection operator formalism \cite{Nakajima1958,Mori1965,Zwanzig1960,MeyerVS2019,teVrugtW2019}, reviewed in Refs.\ \cite{teVrugtW2019d,Grabert1982,Schilling2021,KlipensteinTJSvdV2021}. This formalism allows to obtain transport equations by projecting the microscopic dynamics onto an arbitrary set of \ZT{relevant variables}. The Mori-Zwanzig formalism has been applied successfully in a variety of contexts including fluid dynamics \cite{Grabert1982,CamargodlTDZEDBC2018} (also at surfaces or interfaces \cite{BocquetB1994,CamargodlTDBCE2019,BausT1983}), DDFT \cite{Yoshimori2005,EspanolL2009}, extensions of DDFT \cite{WittkowskiLB2012,WittkowskiLB2013,AneroET2013}, and general relativity \cite{teVrugtHW2021}. Moreover, it has a natural connection to irreversible dynamics \cite{teVrugt2022}, in particular nonequilibrium thermodynamics \cite{teVrugtW2019d}. Consequently, it is a very promising approach for a microscopic derivation of thin-film hydrodynamics. 

In this article, we present a microscopic derivation of the thin-film equation using the Mori-Zwanzig formalism. The dynamics of the fluid particles is projected onto the film height, which is chosen as a relevant variable. Thereby, the thin film equation is obtained almost directly, along with microscopic expressions for the mobility and the free energy functional, and without a need for the full hydrodynamic theory. Our results are verified against molecular dynamics and continuum simulations. The formalism provides a natural route to the derivation of extensions of the standard thin film equation.

This article is structured as follows: In \cref{equations}, we introduce the governing equations. An introduction to the Mori-Zwanzig formalism is provided in \cref{mzformalism}. The microscopic description is developed in \cref{micro}. In \cref{derivation}, we derive the thin-film equation. We explain how to obtain the free energy functional in \cref{freeenergy}. An extension of the thin-film equation with memory is developed in \cref{memory}. In \cref{standard}, we compare our results to standard derivations. Simulations are presented in \cref{simulation}. We conclude in \cref{conclusion}.

\section{\label{equations}Governing equations}
Our aim is the microscopic derivation of the thin film equation
\begin{equation}
\partial_t h\rt= \Nabla \cdot \Big(Q(h\rt)\Nabla \Fdif{F}{h\rt}\Big).
\label{governingequation}
\end{equation}
describing the time evolution of the film height $h$ as a function of (two-dimensional) position $\vec{r}$ and time $t$. Here, $Q$ is the mobility and $F$ is the free energy. Typically, one considers the case $Q =h^3/(3\eta)$ with the dynamic viscosity $\eta$, which corresponds to no-slip boundary conditions. However, other cases are also possible, such as $Q \propto h^2$ corresponding to strong slip. Equation \eqref{governingequation} has the form of a gradient dynamics, which describes the relaxation of a slow conserved variable towards an equilibrium state characterized by a minimum of $F$. (This paragraph follows Ref.\ \cite{HonischLHTG2015}.)

One derives \cref{governingequation} phenomenologically as a limiting case of the incompressible Navier-Stokes equation, as discussed in detail in Ref.\ \cite{Engelnkemper2017}. The derivation, which is rather involved, consists of imposing certain boundary conditions (no-slip at the substrate, force balance and kinematic boundary condition at the free surface) and making a long-wave approximation corresponding to the assumption that horizontal length scales are much longer than vertical length scales. In this work, we will use a different approach based on the Mori-Zwanzig formalism. This method allows to derive equations of the form \eqref{governingequation} directly from the microscopic dynamics.

\section{\label{mzformalism}Mori-Zwanzig projection operator formalism}
The Mori-Zwanzig formalism \cite{Mori1965,Zwanzig1960,Nakajima1958} is a method of nonequilibrium statistical mechanics that allows for the microscopic derivation of transport equations for an arbitrary set of \textit{relevant variables} from the microscopic dynamics of a many-particle system. A general introduction to this formalism can be found in Refs.\ \cite{Grabert1982,teVrugtW2019d}. We here describe it loosely following Ref.\ \cite{WittkowskiLB2013}.

Suppose that we wish to describe a system of $N$ particles with positions $\vec{r}_i$ and momenta $\vec{p}_i$, governed by a Hamiltonian $H$, that is microscopically described by the phase-space distribution $\rho$ that, in general, is unknown. All that we know about the system are the mean values $\{a_i\}$ of a set of $\kappa$ macroscopic observables $\{A_i\}$ that are defined on phase space. Therefore, we approximate the unknown distribution $\rho$ in terms of a \ZT{relevant} distribution $\bar{\rho}$ that depends only on the relevant variables. Assuming maximal informational entropy with respect to the unknown degrees of freedom, the relevant density has the form
\begin{equation}
\bar{\rho}(t)=\frac{1}{\Xi(t)}\exp\bigg(-\beta (H - \mu N -\sum_{i=1}^{\kappa}a^\natural_j(t) A_j(t))\bigg)
\end{equation}
with the grand-canonical partition function $\Xi$, the thermodynamic beta $\beta= (k_\mathrm{B} T)^{-1}$ with Boltzmann constant $k_\mathrm{B}$ and temperature $T$, the chemical potential $\mu$, the particle number $N$, and the thermodynamic conjugates $\{a_j^\natural\}$ that are chosen in such a way that the conditions
\begin{align}
\Tr(\bar{\rho}(t))&=1,\\
\Tr(\bar{\rho}(t)A_i) &=a_i(t)
\end{align}
are satisfied. The trace $\Tr$ is, for a grand-canonical system in three dimensions, given by \cite{teVrugtLW2020}
\begin{equation}
\Tr(Y) = \sum_{N=0}^{\infty}\frac{1}{N!(2\pi\hbar)^{3N}}\INT{}{}{^3 r_1}\INT{}{}{^3 p_1}\dotsb\INT{}{}{^3 r_N}\INT{}{}{^3 p_N}Y,
\end{equation}
where $Y$ an arbitrary function and $\hbar$ is the reduced Planck constant. If we introduce the free energy functional
\begin{equation}
F = \Tr(\bar{\rho}H) + k_\mathrm{B} T \Tr(\bar{\rho}\ln(\rho)),  
\label{freeenergyfunctional}
\end{equation}
the thermodynamic conjugates can be expressed as
\begin{equation}
a^\natural_j(t)=\pdif{F}{a_j(t)}. 
\label{conjugate}
\end{equation}

We are now interested in the dynamics of the mean values $\{a_j\}$. Microscopically, one can infer from Hamilton's equations that the variables $\{A_i\}$ satisfy
\begin{equation}
\dot{A}_i = \ii L A_i 
\end{equation}
with the Liouvillian $\ii L$ that is defined as
\begin{equation}
\ii L = \sum_{i=1}^{N}(\Nabla_{\vec{p}_i}H)\cdot\Nabla_{\vec{r}_i} - (\Nabla_{\vec{r}_i}H)\cdot\Nabla_{\vec{p}_i}. 
\end{equation}
For describing the macroscopic dynamics, one introduces a projection operator $\mathcal{P}$, defined as
\begin{equation}
\mathcal{P}(t)Y = \Tr(\bar{\rho}(t)Y) + \sum_{j=1}^{\kappa}(A_j - a_j(t))\pdif{}{a_j(t)}\Tr(\bar{\rho}(t)Y),
\label{projectionoperator}
\end{equation}
that projects the full microscopic dynamics onto the closed subdynamics of the relevant variables. Moreover, one introduces a complementary projection operator $\mathcal{Q} = 1-\mathcal{P}$. 

The time evolution of the mean values of the relevant variables is then given by \cite{teVrugtLW2020}
\begin{equation}
\dot{a}_i(t) = \organized_i(t) - \sum_{j=1}^{\kappa}\INT{0}{t}{s}R_{ij}(t,s)\beta a^\natural_j(s) + \randomf_i(t,0)
\label{mzexact}
\end{equation}
with the organized drift
\begin{equation}
\organized_i(t) = \Tr(\bar{\rho}(t)\ii L A_i),
\end{equation}
the retardation matrix
\begin{equation}
R_{ij}(t,s)=\Tr(\bar{\rho}(s)(\mathcal{Q}(s)\mathcal{G}(s,t)\ii LA_i)\ii LA_j),
\label{retardationmatrix}
\end{equation}
the orthogonal dynamics propagator
\begin{equation}
\mathcal{G}(s,t)=\exp_\mathrm{R}\!\bigg(\TINT{s}{t}{t'}\ii L\mathcal{Q}(t')\bigg),
\end{equation}
the mean random force
\begin{equation}
\randomf_i(t,0)=\braket{\mathcal{Q}(0)\mathcal{G}(0,t)\ii LA_i} 
\label{eq:meanrandomforce}%
\end{equation}
with the ensemble average $\braket{\cdot}$, and the right-time-ordered exponential $\exp_\mathrm{R}(\cdot)$.

We now assume that
\begin{enumerate}
    \item the macroscopic variables provide a full description of the macroscopic dynamics in the sense that all other variables relax quickly,
    \item $\rho(0)=\bar{\rho}(0)$, i.e, that the system starts in a state of constrained equilibrium.
\end{enumerate}

In this case, one can show \cite{teVrugtW2019} that \cref{mzexact} can be approximated by
\begin{equation}
\dot{a}_i(t) = \organized_i(t) - \sum_{j=1}^{\kappa}D_{ij}(t)\beta a^\natural_j(t)
\label{markov}
\end{equation}
with the diffusion tensor
\begin{equation}
D_{ij}(t) = \INT{0}{\infty}{s}\Tr(\bar{\rho}(t)(\mathcal{Q}(t)\ii L A_j)e^{\ii L s}(\mathcal{Q}(t)\ii L A_i)).  
\label{diffusiontensor}
\end{equation}
A violation of assumption 1 leads to transport equations that are non-local in time (memory effects), a violation of assumption 2 leads to transport equations that contain a mean random force term. It is easily shown that, if the dynamic is given by \cref{markov}, the free energy $F$ given by \cref{freeenergyfunctional} is monotonically decreasing. If there is only one relevant variable, the organized drift typically vanishes for symmetry reasons.

In this work, the relevant variable is a conserved field $A$. For conserved fields in $d$ spatial dimensions, we can introduce a current $\vec{J}\rt$ defined by
\begin{equation}
\ii L A_i\rt = - \Nabla \cdot \vec{J}_i\rt.
\end{equation}
This allows to rewrite \cref{markov} as
\begin{equation}
\begin{split}
\pdif{}{t}a_i\rt &= - \Nabla_{\vec{r}} \cdot \Tr(\bar{\rho}(t)\vec{J}_i(\vec{r},0))\\
&\quad\, + \sum_{j=1}^{\kappa}\Nabla_{\vec{r}}\cdot \bigg(\INT{}{}{^d r}D_{ij}(\vec{r},\vec{r}',t)\beta\Nabla_{\vec{r}'}a_j^\natural(\vec{r}',t)\bigg)
\label{eddftequation}
\end{split}    
\end{equation}
with the diffusion tensor
\begin{equation}
D_{ij}(\vec{r},\vec{r}',t)=\INT{0}{\infty}{s}\Tr(\bar{\rho}(t)(\mathcal{Q}(t)\vec{J}_j(\vec{r}',0))e^{\ii L s}\otimes(\mathcal{Q}(t)\vec{J}_i(\vec{r},0)), 
\label{generaldiffusiontensor}
\end{equation}
where $\otimes$ is the dyadic product. Note that \cref{conjugate} changes to
\begin{equation}
a_j^\natural\rt = \Fdif{F}{a_j\rt},    
\end{equation}
i.e., we now have a functional rather than an ordinary derivative \cite{teVrugtW2019d}. As can be seen, the derivation in the projection operator framework naturally leads to a gradient dynamics form for the dissipative part. Moreover, it can be shown that \cref{eddftequation} allows to prove an H-theorem for the free energy functional \cite{AneroET2013,teVrugt2022}.

\section{\label{micro}Microscopic dynamics and relevant variables}
We consider a system of $N$ particles with mass $m$, where $\vec{r}_i\dd$ is the position and $\vec{p}_i\dd$ the momentum of particle $i$. (Following Ref.\ \cite{Engelnkemper2017}, we will from now on use a superscript $\dd$ to denote three-dimensional vectors, all other vectors are assumed to be two-dimensional. For example, we have $\vec{r}\dd = (x,y,z)^\mathrm{T}$ and $\vec{r}=(x,y)^\mathrm{T}$.) The Hamiltonian is given by
\begin{align}
H&=\sum_{i=1}^{N}H_i,\\
H_i&=\frac{(\vec{p}_i\dd)^2}{2m} + \frac{1}{2}\sum_{j\neq i}^{N}U_2(\vec{r}_i\dd - \vec{r}_j\dd)+U_1(\vec{r}_i\dd).
\end{align}
Here, $U_2$ is the interaction potential and $U_1$ is the external potential, which we here assume to be time-independent.

The first step is the microscopic definition of the observable of interest, in our case the film height. We choose
\begin{equation}
\hat{h}(\vec{r})=\vorf\INT{}{}{z}\hat{\rho}(\vec{r}\dd)  
\label{filmheight}
\end{equation}
with the density of the liquid phase $\rho_\mathrm{l}$ and the density operator \cite{Dean1996}
\begin{equation}
\hat{\rho}(\vec{r}\dd) = \sum_{i=1}^{N}\delta(\vec{r}\dd - \vec{r}_i\dd),    
\end{equation}
where $\vec{r}\dd_i$ the position of particle $i$. The definition \eqref{filmheight} is adapted from Eq. (7) in Ref.\ \cite{HughesTA2015} (we here assume the density of the gas phase to be negligible compared to the density of the liquid phase). Obviously, the definition \eqref{filmheight} only makes physical sense if the particles are located in a film at a surface, which is what we assume throughout this derivation. Formally, it is of course also applicable otherwise, although in this case $h$ should not be thought of as a film height (and most of our approximations will not be justified).

The microscopic rate of change is found to be
\begin{equation}
\begin{split}
\ii L \hat{h}\rt &= \vorf\Nabla_{\vec{r}_i\dd}\cdot\INT{}{}{z}\sum_{i=1}^{N}\vec{v}_i\dd\delta(\vec{r}\dd - \vec{r}_i\dd(t))\\
&= -\vorf\Nabla_{\vec{r}}\cdot\INT{}{}{z}\sum_{i=1}^{N}\vec{v}_i\delta(\vec{r}\dd- \vec{r}_i\dd(t)),
\end{split}
\end{equation}
where $\vec{v}_i\dd = \vec{p}_i\dd/m$ is the velocity of the $i$-th particle and $\vec{v}_i$ the vector containing the $x$- and $y$-component of $\vec{v}_i\dd$. We have exploited that, since the third coordinate is integrated over, the partial derivative with respect to this coordinate vanishes. Thus, despite the fact that the microscopic dynamics is three-dimensional, we can use a two-dimensional current. The time evolution of $\hat{h}$ can therefore be expressed as
\begin{equation}
\partial_t \hat{h}\rt = - \Nabla \cdot \hat{\vec{J}}\rt    
\end{equation}
with the two-dimensional microscopic current
\begin{equation}
\hat{\vec{J}}\rt = \vorf \sum_{i=1}^{N}\INT{}{}{z}\vec{v}_i\delta(\vec{r}\dd- \vec{r}_i\dd(t)).   
\label{twodcurrent}
\end{equation}
We write $\Nabla$ for $\Nabla_{\vec{r}}$.
\section{\label{derivation}Derivation of transport equations}
For the current \eqref{twodcurrent}, the organized drift term in \cref{eddftequation} vanishes for symmetry reasons (integral over an odd function of the momenta). Thus, we find
\begin{equation}
\partial_t h \rt = \Nabla \cdot \INT{}{}{^2r'}\beta D(\vec{r},\vec{r}',t)\Nabla'\Fdif{F}{h(\vec{r}',t)}
\label{nonlocaltf}
\end{equation}
with $\Nabla'=\Nabla_{\vec{r}'}$. The diffusion tensor is given by
\begin{equation}
D(\vec{r},\vec{r}',t) = \INT{0}{\infty}{s}\Tr\!\big(\bar{\rho}(t)\hat{\vec{J}}(\vec{r}',0)\otimes\hat{\vec{J}}(\vec{r},s)\big),  
\label{diffusiontensor1}
\end{equation}
having exploited that\footnote{The trace $\Tr$ includes an integral over the momenta $\vec{p}_i$. We have $\Tr(\bar{\rho}\vec{J}) = \vec{0}$ since $\bar{\rho}$ is even and $\vec{J}$ is odd in the momenta. By \cref{projectionoperator}, this implies $\mathcal{P} \vec{J} = \vec{0}$, which gives $\mathcal{Q}\vec{J} = (1-\mathcal{P})\vec{J} = \vec{J}$.} $\mathcal{Q}\vec{J} = \vec{J}$.
Since the relevant density $\bar{\rho}$ appearing in \cref{diffusiontensor1} for the diffusion tensor is still difficult to work with, we can assume that we are near equilibrium and can therefore replace it by the equilibrium distribution $\rho_{\mathrm{eq}}$ \cite{EspanolL2009} for the purposes of numerical simulations. Basically, the average in \cref{diffusiontensor1} is a  \ZT{constrained equilibrium ensemble average} where \ZT{constrained equilibrium} means that the film height is constrained to be $h\rt$.

Using \cref{twodcurrent}, we can write \cref{diffusiontensor1} as
\begin{equation}
D(\vec{r},\vec{r}',t) = \INT{}{}{z}\INT{}{}{z'} D_z(\vec{r},z,\vec{r}',z',t) \label{drr}.
\end{equation}
We have introduced here the extended diffusion tensor
\begin{equation}
D_z(\vec{r},z,\vec{r}',z',t) = \INT{0}{\infty}{s}\Tr\!\big(\bar{\rho}(t)\hat{\vec{J}}_z(\vec{r}',z',0)\otimes\hat{\vec{J}}_z(\vec{r},z,s)\big) 
\label{diffusiontensorb}
\end{equation}    
with 
\begin{equation}
\hat{\vec{J}}_z(\vec{r}\dd,t) = \vorf \sum_{i=1}^{N}\vec{v}_i\delta(\vec{r}\dd- \vec{r}_i\dd(t)).    
\end{equation}
Assuming translational and rotational invariance in the $xy$-plane, we can write \cref{diffusiontensorb} as
\begin{equation}
D_z(\vec{r},z,\vec{r}',z',t) = \frac{1}{2}\INT{0}{\infty}{s}\braket{\hat{\vec{J}}_z(\vec{0},z',0)\cdot\hat{\vec{J}}_z(\vec{r}-\vec{r}',z,s)}_t\Eins
\label{diffusiontensor2}
\end{equation} 
with the two-dimensional unit matrix $\Eins$, where we have introduced the notation
\begin{equation}
\braket{Y}_t = \Tr(\bar{\rho}(t)Y)
\end{equation}
with an arbitrary phase-space variable $Y$ to emphasize that we are dealing with constrained equilibrium averages. Essentially, $\braket{Y}_t$ is an average over an ensemble of systems that all have a certain film height $h(\vec{r},t)$.

We now perform a Fourier transformation, giving
\begin{equation}
\begin{split}
&D_z(\vec{r},z,\vec{r}',z',t) \\
&= \frac{1}{2\pi}\INT{}{}{^2 k}e^{-\ii \vec{k}\cdot (\vec{r}-\vec{r}')}\tilde{D}(\vec{k},z,z',t)
\label{fouriertransformation}
\end{split}
\end{equation}
with the partially Fourier-transformed extended diffusion tensor $\tilde{D}$. In a thin film, gradients in the $xy$ direction are small compared to gradients in the $z$ direction. If gradients in the $xy$ direction are small, we can set $\vec{k}$ to zero and find
\begin{equation}
\tilde{D}(\vec{k},z,z',t) \approx \tilde{D}(\vec{0},z,z',t).
\label{twoapproximations}
\end{equation}
Inserting \cref{twoapproximations} into \cref{fouriertransformation} gives 
\begin{equation}
\begin{split}
&D_z(\vec{r},z,\vec{r}',z',t) \\
&= \frac{1}{2\pi}\INT{}{}{^2 k}e^{-\ii \vec{k}\cdot (\vec{r}-\vec{r}')}\tilde{D}(\vec{0},z,z',t)\\
&=2\pi\tilde{D}(\vec{0},z,z',t)\delta(\vec{r}-\vec{r}').
\end{split}
\label{approximated}
\end{equation}
The tensor $\tilde{D}(\vec{0},z,z',t)$ is given by
\begin{equation}
\tilde{D}(\vec{0},z,z',t)=\frac{1}{4\pi}\INT{}{}{^2r}\INT{0}{\infty}{s}\braket{\hat{\vec{J}}_z(\vec{r},z,s)\cdot \hat{\vec{J}}_z(\vec{0},z',0)}_t\Eins.
\label{correctdefinition}
\end{equation}
Next, we introduce the \textit{transverse current} as\footnote{Compared to the usual definition, we add a prefactor $1/\rho_{\mathrm{l}}$ in the transverse current and a prefactor $1/\rho_{\mathrm{l}}^2$ in the transverse current correlation for convenience.} \cite{Palmer1994,BocquetB1994}
\begin{equation}
\hat{\vec{J}}_\mathrm{T}(z,t)=\vorf\sum_{i=1}^{N}\vec{v}_i\delta(z-z_i(t)) = \INT{}{}{^2r}\hat{\vec{J}}_z(\vec{r}\dd,t),
\label{eq_J_T}
\end{equation}
where $z_i$ is the third component of $\vec{r}_i\dd$. The \textit{transverse current correlation} is then defined as
\begin{equation}
\begin{split}
C(z,z',t,s) &= \frac{1}{2}\braket{\hat{\vec{J}}_\mathrm{T}(z,s)\cdot\hat{\vec{J}}_\mathrm{T}(z',0)}_t\\
&=\INT{}{}{^2r}\INT{}{}{^2r'}\braket{\hat{\vec{J}}_z(\vec{r},z,s)\cdot\hat{\vec{J}}_z(\vec{r}',z',0)}_t,
\end{split}
\label{definitionofc}
\end{equation}
where we have used a constrained rather than an equilibrium ensemble average. Exploiting translational invariance, \cref{definitionofc} can be re-written as
\begin{equation}
\begin{split}
C(z,z',t,s) &=\frac{1}{2}\INT{}{}{^2r}\INT{}{}{^2r'}\braket{\hat{\vec{J}}_z(\vec{r}-\vec{r}',z,s)\cdot\hat{\vec{J}}_z(\vec{0},z',0)}_t\\
&=\frac{A}{2}\INT{}{}{^2r}\braket{\hat{\vec{J}}_z(\vec{r},z,t)\cdot\hat{\vec{J}}_z(\vec{0},z',0)}_t,
\end{split}
\label{definitionofc2}
\end{equation}
where $A = \TINT{}{}{^2r}$ is the area of the two-dimensional domain. Comparing \cref{correctdefinition,definitionofc2} shows that
\begin{equation}
\tilde{D}(\vec{0},z,z',t)=\frac{1}{2\pi A}\INT{0}{\infty}{s}C(z,z',t,s)\Eins.
\label{cdrelation}
\end{equation}
Inserting \cref{cdrelation} into \cref{approximated} yields
\begin{equation}
D_z(\vec{r},z,\vec{r}',z',t)=\frac{1}{A}\delta(\vec{r}-\vec{r}')\INT{0}{\infty}{s}C(z,z',s)\Eins.
\label{approximated3}
\end{equation}

As is well known \cite{RopoAJ2016,Hess2002,Palmer1994,JakseP2013}, the transverse current correlation function can be related to the viscosity. It obeys the diffusion equation \cite{BocquetB1994}
\begin{equation}
\partial_s C(z,z',t,s)= \nu\partial_z^2 C(z,z',t,s)
\label{diffusionequation}
\end{equation}
with the kinematic viscosity 
\begin{equation}
\nu = \frac{\eta}{m \rho_{\mathrm{l}}}.
\label{kinematicviscosity}
\end{equation}
The dependence on time $t$ is not relevant for the dynamics of the correlation functions. Equation \eqref{diffusionequation} is a standard result, a derivation can be found in Refs.\ \cite{HansenMD2009,Forster1989}. It can be used in two different ways. Since it establishes a relation between the correlation function $C$ and the viscosity $\eta$, one can use it (a) for determining the viscosity $\eta$ from a given correlation function $C$, and (b) for calculating the correlation function $C$ for a given value of $\eta$. Route (a) would be closer to the way in which the Mori-Zwanzig formalism is traditionally applied, namely as a way of obtaining microscopic expressions for transport coefficients. Using the Fourier-transformed transverse current correlation function\footnote{Typically, one sets $z=z'$ such that $C$ depends only on one spatial coordinate, and consequently only on one wavenumber $k_z$.} $\tilde{C}(k_z,\omega)$ with wavenumber $k_z$ and frequency $\omega$, the viscosity is given by \cite{Palmer1994}
\begin{equation}
\lim_{\omega \to 0}\lim_{k_z \to 0}k_\mathrm{B} T\frac{\omega^2}{k_z^2}\tilde{C}(k_z,\omega) = \eta.
\end{equation}
Microscopically, such a connection between the transverse correlation function and the viscosity is a consequence of linear response theory \cite{Forster1989,Palmer1994}. However, for liquids commonly considered in thin-film physics (such as water), the viscosity is well studied and typically known. What is far more interesting in practice is to thus use \cref{diffusionequation} to calculate, for a known value of $\eta$, the correlation function $C$. As will be shown, this allows us to calculate the mobility $Q(h)$ from the thin-film equation from first principles.

Equation \eqref{diffusionequation} can be obtained by assuming, following the line of argument in Ref.\ \cite{BocquetB1994}, that the equilibrium fluctuations of the momentum density obey the same evolution equations and boundary conditions as the momentum density itself (a generalization of Onsager's linear regression hypothesis \cite{Onsager1931,Onsager1931b}). The macroscopic velocity field obeys the boundary condition \cite{Engelnkemper2017}

\begin{equation}
\vec{v}|_{z=0} =\vec{0}\label{bca}
\end{equation}
at the surface (no-slip condition). From \cref{bca}, it then follows that $C$ obeys
\begin{equation}
C(z,z',t,s)|_{z=0} =0\label{bca1}. 
\end{equation}
From equilibrium statistical mechanics, one can derive the initial condition \cite{BocquetB1994}
\begin{equation}
C(z,z',t,0)= \frac{k_\mathrm{B} T}{m\rho_{\mathrm{l}}^2}\braket{\hat{\rho}_z}_t\delta(z-z')  
\label{initialconditionb2}
\end{equation}
with the $xy$-averaged particle density
\begin{equation}
\hat{\rho}_z = \sum_{i=1}^{N}\delta(z-z_i)   
\end{equation}
in the canonical ensemble. Making the plausible assumption that $\braket{\hat{\rho}_z}_t = \rho_{\mathrm{l}}A$ for $0< z < h$, \cref{initialconditionb2} becomes
\begin{equation}
C(z,z',t,0)= \frac{k_\mathrm{B} T A}{m \rho_{\mathrm{l}}}\delta(z-z')
\label{bbinitialcondition}
\end{equation}
for $0 < z < h$. The result \eqref{initialconditionb2} can be obtained from (canonical) equilibrium statistical mechanics (see Ref.\ \cite{BocquetB1994}). Since $h$ is a slow variable, we may assume that it is constant for the purposes of calculating the mobility function. To reduce the number of prefactors, we consider the integrated correlation function 
\begin{equation}
M(z,t,s)=\TINT{}{}{z'}\frac{m \rho_{\mathrm{l}}}{k_\mathrm{B} T A} C(z,z',t,s) 
\label{approximated2}
\end{equation}
instead. Thus, we wish to solve the differential equation
\begin{equation}
\partial_s M(z,t,s)= \nu\partial_z^2 M(z,t,s)
\label{diffusionequationb}
\end{equation}
for $z \in [0,\infty)$ with the initial condition
\begin{equation}
M(z,t,0)= 
\begin{cases}
1 &\text{ for } 0<z<h,\\
0 &\text{ otherwise}
\end{cases}
\label{initialcondition}
\end{equation}
and the boundary condition 
\begin{equation}
M(0,t,s)=0. 
\label{boundarycondition}
\end{equation}
Using standard techniques \cite{Selvadurai2000,Widder1975}, the solution of \cref{diffusionequationb} for the initial condition \eqref{initialcondition} and the boundary condition \eqref{boundarycondition} is found to be
\begin{widetext}
\begin{equation}
M(z,t,s)=\frac{1}{2}\bigg(2\erf\!\bigg({\frac{z}{\sqrt{4\nu s}}}\bigg) - \erf\!\bigg({\frac{z-h_t}{\sqrt{4\nu s}}}\bigg) - \erf\!\bigg({\frac{z+h_t}{\sqrt{4\nu s}}}\bigg)\bigg)
\label{mzt}
\end{equation}
\end{widetext}
with the error function $\erf{}(\cdot)$ defined as 
\begin{equation}
\erf(x)=\frac{2}{\sqrt{\pi}}\INT{0}{x}{t}e^{-t^2}. 
\end{equation}
We have briefly written $h_t = h(\vec{r},t)$ (and ignored the dependence of $M$ on $\vec{r}$). Of course, \cref{mzt} will in practice only be valid for $z \in (0,h)$, since there are no particles in the substrate or above the film. We require the time integral over \cref{mzt}, which is given by
\begin{equation}
\tilde{M}(z,t)=\INT{0}{\infty}{s}M(z,t,s)=\frac{1}{\nu}\bigg(zh_t-\frac{z^2}{2}\bigg).
\label{timeintegral}
\end{equation}
Combining \cref{drr,approximated3,kinematicviscosity,approximated2,timeintegral} and dropping the subscript $t$, we find
\begin{equation}
\begin{split}
D(\vec{r},\vec{r}',t)&=\INT{0}{h}{z}\frac{k_\mathrm{B} T}{\eta}\bigg(zh-\frac{z^2}{2}\bigg)\delta(\vec{r}-\vec{r}')\Eins\\ 
&= k_\mathrm{B} T\frac{h^3}{3\eta}\delta(\vec{r}-\vec{r}')\Eins,
\end{split}
\label{dvonrrp}
\end{equation}
where we have assumed that the integral over $z$ in \cref{drr} can be restricted to the interval $[0,h]$ (for $z > h$, there are no particles, such that the correlation function has to vanish). Finally, inserting \cref{dvonrrp} into \cref{nonlocaltf}, we find
\begin{equation}
\partial_t h = \Nabla \cdot \bigg(\frac{h^3}{3\eta}\Nabla \Fdif{F}{h}\bigg),  
\label{governingequationspec}
\end{equation}
which is the thin-film equation \eqref{governingequation}.

\section{\label{freeenergy}Free energy functional}
What is left is the calculation of the free energy functional $F$. This, however, turns out to be surprisingly simple, since we can make use of the \textit{bridge theorem} derived by \citet{AneroET2013}. It states that if we have two levels of description A and B, where A is a coarse-grained and B a fine-grained description, and if the relevant variables of A are linear functions of the relevant variables of B, then the entropy functional of level A is obtained by maximizing the entropy of level B subject to the constraint that the average of the relevant variables of B gives the relevant variables of A. 

In our case, level A is given by thin-film hydrodynamics, and level B by dynamical density functional theory \cite{teVrugtLW2020}. The film height $h$ is -- in our definition -- a linear function of the density $\rho$. Thus, we can obtain the entropy $S[h]$ by maximizing the entropy $S[\rho]$ subject to the constraint $h = \vorf\TINT{}{}{z}\rho$. Due to the simple relation $F[\rho] = \braket{H} - TS[\rho]$ with the ensemble average of the Hamiltonian $\braket{H}$, this can be directly translated into a minimization principle for the free energy:
\begin{equation}
F[h] = \underset{\vorf\TINT{}{}{z}\rho = h}{\min} (F_{\mathrm{DFT}}[\rho]).
\label{abstract}
\end{equation}
Thus, we know the free energy functional $F[h]$ once we have a suitable DFT functional $F_{\mathrm{DFT}}$. 

In practice, there is still some work to be done to come from the abstract definition \eqref{abstract} to a directly applicable formula for the free energy $F[h]$. Fortunately, this has already been done by other authors \cite{HughesTA2015,HughesTA2017,ArcherE2011}, such that we can restrict ourselves here to a brief presentation of what needs to be done. We follow Ref.\ \cite{HughesTA2017}.

The grand-canonical potential is given by
\begin{equation}
\Omega = - pV + A(\gamma_{\mathrm{wl}}+ \gamma_{\mathrm{lg}} + g(h) + \Gamma \delta\mu)   
\label{omega}
\end{equation}
with pressure $p$, volume $V$, area $A$, wall-liquid interfacial tension $\gamma_{\mathrm{wl}}$, liquid-gas interfacial tension $\gamma_{\mathrm{lg}}$, adsorption $\Gamma$, and deviation from chemical potential at coexistence $\delta \mu$. The most interesting quantity is the binding potential $g(h)$, which is defined by \cref{omega}. It is related to the disjoining pressure $\Pi$ as
\begin{equation}
\Pi = -\pdif{g}{h}.    
\end{equation}
Microscopically, $\Omega$ can be calculated in DFT as
\begin{equation}
\begin{split}
\Omega &= k_\mathrm{B} T \INT{}{}{^3r}\rho(\vec{r})(\ln(\Lambda^3\rho(\vec{r})) -1) \\
&\quad\,+F_{\mathrm{exc}} + \INT{}{}{^3r}\rho(U_1(\vec{r}) - \mu)
\end{split}
\end{equation}
with the thermal de Broglie wavelength $\Lambda$, the excess free energy $F_{\mathrm{exc}}$, the chemical potential $\mu$. 

For the purpose of describing thin films, only terms that are at least quadratic in $h$ are relevant. This gives
\begin{equation}
F[h]= \INT{}{}{^2r} \Big( g(h) + \gamma_{\mathrm{lg}}\sqrt{1+(\Nabla h)^2} \Big).
\end{equation}
The first term in the integrand is the binding potential, the second term takes into account that the surface area depends on the film height. It is often approximated as $\frac{\gamma_{\mathrm{lg}}}{2}(\Nabla h)^2$, a form that arises from a Taylor expansion of the square root with dropping a constant term. Thus, we need to calculate $\gamma_{\mathrm{lg}}$ and $g(h)$ from DFT. For obtaining $\gamma_{\mathrm{lg}}$, one calculates the density profile of the gas-liquid interface for a vanishing external potential, uses this profile to obtain the free energy $\Omega_{\mathrm{lg}}$, and then finds
\begin{equation}
\gamma_{\mathrm{lg}} = \frac{\Omega_{\mathrm{lg}} + pV}{A}.  
\end{equation}
Obtaining the binding potential is more difficult. For this purpose, one introduces into the free energy a fictitious external potential constraining the system to a certain film height and thereby calculates the free energy for various film heights. Then, one makes an ansatz for the function $g(h)$ (see Eqs.\ (26) and (27) in Ref.\ \cite{HughesTA2017}) and fits the free parameters to the simulation results.

\section{\label{memory}Thin-film equation with memory effects}
\begin{figure*}
\centering\includegraphics[width=\linewidth]{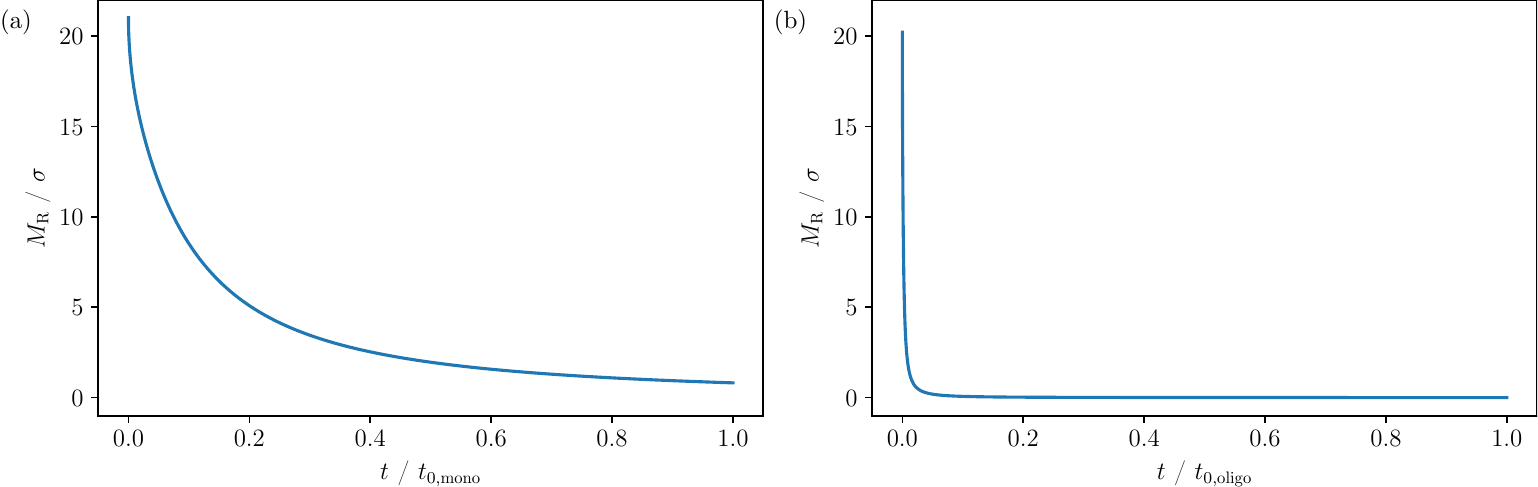}
\caption{\label{fig:1} Plot of the function $M_\mathrm{R}(h_t,t)$ given by \cref{mr} for (a) monomers ($\rho_\mathrm{l}=0.76\sigma^{-3}$, $h_t = 21.02\sigma$, $\nu = 2.41\sigma^2/\tau$) and (b) oligomers ($\rho_\mathrm{l}=1.06\sigma^{-3}$, $h_t = 20.22\sigma$, $\nu = 23.4\sigma^2/\tau$). The time $t$ is specified in units of the characteristic timescale for the thin film equation, which has the values $t_{0,\mathrm{mono}}=241.82\tau$ for monomers  in (a) and $t_{0,\mathrm{oligo}} = 1275.09\tau$ in (b).}
\end{figure*}
As discussed in \cref{mzformalism}, dissipative Markovian transport equations such as \cref{governingequation} are, in general, only an approximation to the exact dynamics which actually also depends on previous times. More precisely, the thin-film equation \eqref{governingequation} has been shown to be a special case of the general transport equation \eqref{markov}, which in turn is an approximation to \cref{mzexact}. Such approximations, while often very accurate, have certain drawbacks. An example known from DDFT is the overestimation of relaxation times \cite{teVrugtLW2020}.

Therefore, it would be interesting to derive a generalization of the thin-film equation that corresponds to \cref{mzexact} rather than \cref{markov}, i.e., that takes into account also time-delay. To this end, we note that, if we replace the orthogonal dynamics propagator $\mathcal{G}(s,t)$ in \cref{retardationmatrix} by an ordinary propagator $\exp(\ii L (t-s))$ (a common approximation that is also required for deriving \cref{diffusiontensor} \cite{teVrugtW2019d}), and if we take into account that $\mathcal{Q}(t)\ii L A = \ii L A$ for the variable considered here (see footnote 1), we can write $R_{ij}(t,s)=R_{ij}(s,t-s)$. The relation between the retardation matrix and the diffusion tensor is then given by 
\begin{equation}
D_{ij}(t)=\INT{0}{\infty}{s}R_{ij}(t,s).
\label{timeintegral2}
\end{equation}
See Ref.\ \cite{teVrugtW2019} for a more detailed derivation. If we compare \cref{timeintegral2} to \cref{timeintegral,dvonrrp}, which essentially state that the diffusion tensor is equal to
\begin{equation}
D(\vec{r},\vec{r}',t)=\frac{k_\mathrm{B} T}{m\rho_{\mathrm{l}}}\INT{0}{h_t}{z}\INT{0}{\infty}{s}M(z,t,s)\delta(\vec{r}-\vec{r}')\Eins,
\label{ditensor}
\end{equation}
we can infer that the retardation matrix should be (approximately) equal to
\begin{equation}
R(\vec{r},\vec{r}',t,s)=\frac{k_\mathrm{B} T}{m\rho_{\mathrm{l}}} \INT{0}{h_t}{z}M(z,t,t-s)\delta(\vec{r}-\vec{r}')\Eins.
\label{retardationmatrix2}
\end{equation}
If we insert the function $M(z,t,s)$ given by \cref{mzt} into \cref{retardationmatrix2}, we obtain
\begin{equation}
R(\vec{r},\vec{r}',t,s)=\frac{k_\mathrm{B} T}{m\rho_{\mathrm{l}}} M_\mathrm{R}(h_t,t-s)\delta(\vec{r}-\vec{r}')\Eins
\label{retardationmatrix3}
\end{equation}
with
\begin{widetext}
\begin{equation}
M_\mathrm{R}(h_t,s)=\sqrt{\frac{\nu s}{\pi}}\bigg(4\exp\!\bigg(-\frac{h_t^2}{4\nu s}\bigg) - \exp\!\bigg(-\frac{h_t^2}{\nu s}\bigg) - 3\bigg) + h_t\bigg(2\erf\!\bigg(\frac{h_t}{\sqrt{4\nu s}}\bigg) -\erf\!\bigg(\frac{h_t}{\sqrt{\nu s}}\bigg)\bigg). 
\label{mr}
\end{equation}
\end{widetext}
However, there is one additional difficulty: We need to specify at which point in time the function $h$ appearing in the memory kernel has to be evaluated. In the Markovian limit, we can simply assume it to be constant while calculating the memory kernel (such that we can always use $h\rt$), however, this is no longer true in the non-Markovian case. 

To solve this problem, we can note that in the Markovian limit the relevant density $\bar{\rho}(t)$ in \cref{diffusiontensor1} enforces the value of $h\rt$, i.e., it settles the system to the local equilibrium state corresponding to the time-dependent height profile $h\rt$. In \cref{retardationmatrix}, we have $\bar{\rho}(s)$ instead of $\bar{\rho}(t)$. Consequently, it is reasonable to use $h(\vec{r},s)$ rather than $h\rt$ for the mobility function in the non-Markovian case. Since the $h$ appearing in the thermodynamic force also depends on time $s$, this ensures that every $h$ appearing in the dynamic equation has the same argument.

Hence, by specializing \cref{mzexact} to the case of the relevant variable $h$ and by inserting \cref{retardationmatrix3}, we can infer that a generalization of the thin-film equation that involves memory effects is given by
\begin{equation}
\partial_t h\rt = \INT{0}{t}{s}\Nabla \cdot  \bigg( M_\mathrm{R}(s,t-s)\Nabla \Fdif{F}{h(\vec{r},s)}\bigg).
\label{governingequationmemory}
\end{equation}

To recover the standard thin-film equation \eqref{governingequation}, one then requires the following steps:
\begin{enumerate}
    \item Assume that the relevant variable $h$ is slow. This corresponds to replacing $h(\vec{r},s)$ by $h\rt$ in \cref{governingequationmemory}.
    \item Assume that the memory function decays rapidly for larger values of $t-s$. This allows to change the lower integration boundary in  \cref{governingequationmemory} from 0 to $-\infty$. Finally, we substitute $s \to t-s$ and switch integration boundaries \cite{teVrugtW2019,teVrugtW2019d}.
\end{enumerate}
Due to the time integral and the complicated form \eqref{mr} of $M_\mathrm{R}$, \cref{governingequationmemory} will, in general, be very difficult to deal with even numerically.

\section{\label{standard}Comparison to standard derivation}
The derivation of the thin-film equation presented here has significant advantages compared to the standard derivation from the Navier-Stokes equation:
\begin{itemize}
\item Despite being a microscopic derivation, it is shorter than phenomenological derivations. The transport equation \eqref{governingequation} is obtained pretty directly.

\item It is not necessary to impose kinematic boundary conditions or to make a long-wave approximation, since these are implicitly contained in the choice of $h$ as a relevant variable.

\item We do not need to first consider the full hydrodynamic equations before making approximations.

\item It is easy to generalize \cref{governingequation} to incorporate, e.g., additional fields, nonlocality, or memory effects. We have demonstrated this explicitly for the case of memory in \cref{memory}.

\item Owing to the deep connection between the Mori-Zwanzig formalisn and irreversible thermodynamics \cite{WittkowskiLB2013,teVrugtW2019d}, we get a much better explanation for why the thin-film equation has the gradient dynamics form \eqref{governingequation}.

\item We get microscopic expressions for the mobility and the free energy.
\end{itemize}

A comment on incompressibility is in order: It might be argued that the incompressible Navier-Stokes equation assumes $\rho(\vec{r}\dd)=\rho_{\mathrm{l}}$ = const., such that it is not helpful to define the film height in terms of the density. Actually, however, the incompressible Navier-Stokes equation assumes
\begin{equation}
\rho(\vec{r}\dd) =
\begin{cases}
\rho_{\mathrm{l}} & \text{ for }z \in [0,h],\\
0& \text{ otherwise},
\end{cases}
\end{equation}
such that integrating over $\rho$ does give the film height.

\section{\label{simulation}Molecular dynamics simulations}
\subsection{\label{setup}Simulation setup}
\begin{figure}
    \centering
    \includegraphics[width=\linewidth]{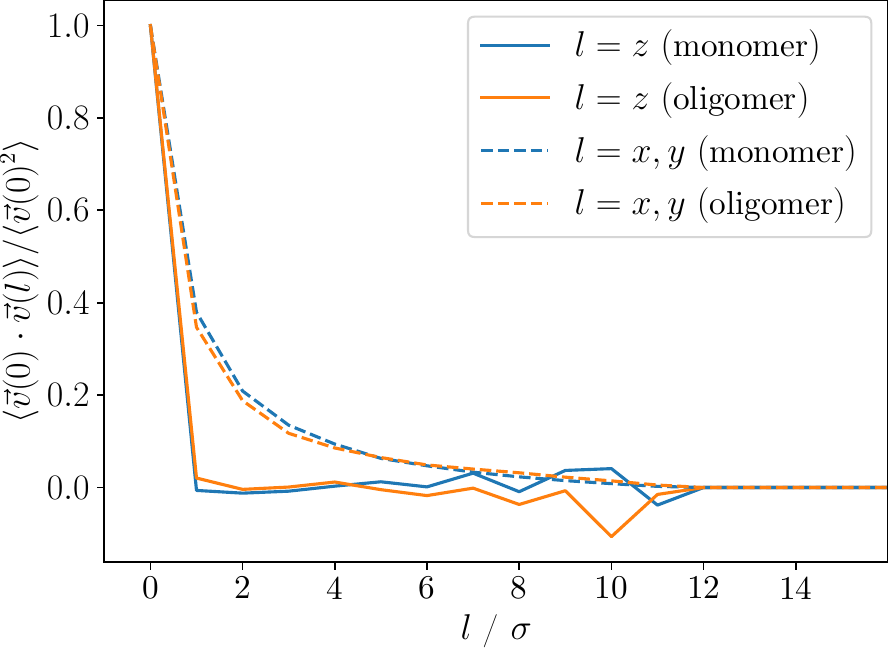}
    \caption{\label{fig:2}Spatial correlation of the velocity $\vec{v}$ in $xy$-direction at time $t_{\mathrm{MD}}= \SI{0}{\tau}$ in $xy$- and $z$-direction for the simple liquid and the oligomer.}
\end{figure}
The theoretical framework developed in this article can be tested using numerical simulations. For this purpose, we can exploit that the derivation based on the Mori-Zwanzig formalism leads to an equation of the form  \eqref{governingequation} along with a microscopic expression for the mobility $Q(h)$. In particular, the derivation presented here makes a specific prediction for the form of the integrated correlation function $M$, which is given by \cref{mzt}. This prediction can be tested by comparing it to the results of Molecular Dynamics (MD) simulations, where the integrated correlation function can be calculated microscopically using \cref{definitionofc,approximated2}.

The MD simulation are performed with the package HOOMD \cite{anderson_hoomd,phillips_hoomd_dpd}. For the choice of the simulation setup, we follow previous work \cite{ToppSGH2022,StienekerTGH2021}. The particles are assumed to interact via the Lennard-Jones potential
\begin{equation}
  V(r_{ij}) = 4 \eps{$ij$} \left( \left(\frac{\sigma}{r_{ij}}\right)^{12} - \left( \frac{\sigma}{r_{ij}} \right)^{6} \right),
  \label{lennardjones}
\end{equation}
with the distance $r_{ij}$ of particles $i$ and $j$, the interaction strength $\eps{lj}$, and the particle diameter $\sigma$. We truncate this potential at a cut-off radius of $r_{c} = \SI{2.5}{\sigma}$. The system contains two particle types (with the same mass and diameter), namely substrate particles ($\text{s}$) whose positions are fixed to two layers of an fcc(111) lattice with lattice constant $a$, and fluid particles ($\text{f}$), which can move. The interaction strengths are given by $\eps{ff}=\eps{sf}=\varepsilon$ with an energy $\varepsilon$ that corresponds to the physical energy scale of the problem. In the simulations, all energies are measured in units of $\varepsilon$, such that the prefactor in \cref{lennardjones} is 4.  

In this work, we set the lattice constant to $a = \SI{1.74}{\sigma}$. By using a total amount of $N = 30,000$ fluid particles inside a simulation box with dimensions $l_{x}=\SI{36.92}{\sigma}$ in $x$- and $l_{y}=\SI{31.97}{\sigma}$ in $y$-direction, the simulation of a liquid film is ensured. The time-step size is chosen to be $t_{\mathrm{MD}} = \SI{0.005}{\tau}$ with $\tau = \sigma\sqrt{m/\varepsilon}$ and the reduced temperature is set to $k_{\text{B}}T/\varepsilon=0.75$.

To check whether our results also apply for polymer films, we also performed MD simulations for oligomers. These are modeled by particles connected in unbranched chains with a finite extensible nonlinear elastic (FENE) potential
\begin{equation}
U_{\mathrm{FENE}}(r) = -\frac{1}{2} k R_{0}^{2} \ln \! \left(1- \left( \frac{r}{R_{0}} \right)^{2} \right),
\end{equation}
where $k = \SI{30}{\varepsilon}$ is the strength of the attractive force and $R_{0} = \SI{1.5}{\sigma}$ the maximum bond length. In this work, we set the chain length to 20 particles. Although our derivation assumes monomers, the results should essentially carry over since the particles in the chains also have a spherically symmetric interaction potential and also obey Hamilton's equations. (The main difference that would make the derivation technically more complicated is that in the oligomer case, the interaction potential between two different particles is not generally the same since they could belong to different chains or be in different positions within the same chain.)

\subsection{Test of theoretical results}
\subsubsection{Validity of Markovian approximation}
The MD simulations described in \cref{setup} allow us to check the accuracy of several assumptions and results made within the analytical derivation.

A very central assumption made in this derivation is the \textit{Markovian approximation}. The thin film equation has, in this work, been found to be a special case of \cref{markov}. This equation, however, is itself only a special case of the more general non-Markovian equation \eqref{mzexact} -- more precisely the special case in which the memory function decays very rapidly (the Markovian approximation is the assumption that this is indeed the case). Using \cref{mzexact} rather than \cref{markov} as a basis for the derivation would lead to the more general equation \eqref{governingequationmemory}, from which the thin film equation arises as a special case only if we assume that the function $M_\mathrm{R}$ decays very rapidly. This is what the Markovian approximation corresponds to in this case.

To judge whether a function decays rapidly, we require a time scale to compare it to. The Markovian approximation corresponds to the assumption of a time scale separation between microscopic and macroscopic degrees of freedom, i.e., it is required that $M_\mathrm{R}$ decays rapidly compared to the time scales on which the film height $h$ evolves. As shown in Ref.\ \cite{StienekerTGH2021}, the characteristic time scale for the thin film equation is $t_0 = 3\eta h_0/\gamma_{\mathrm{lg}}$ with the characteristic film height $h_0$. For the monomers, we have in our simulations $h_0=21.02\sigma$ (typical film height) and $\eta = 1.82919 m/(\sigma\tau)$. The surface tension $\gamma_{\mathrm{lg}}$ has been computed via the anisotropy of the normal and tangential part of the pressure tensor of a liquid slab in its own vapor phase \cite{calc_surf_tension}, similar to the procedure used in Ref.\ \cite{ToppSGH2022}. For the monomer system, we have found $\gamma_{\mathrm{lg,\mathrm{mono}}} = \SI{0.477}{\epsilon/\sigma^2}$, giving $t_{0,\mathrm{mono}} = \SI{241.821219}{\tau}$, and for the oligomer system $\gamma_{\mathrm{lg,\mathrm{oligo}}} = \SI{1.18}{\epsilon/\sigma^2}$ and $t_{0,\mathrm{oligo}} = \SI{1275.09376}{\tau}$

Figure \ref{fig:1} shows a plot of the function $M_\mathrm{R}(h_t,t)$ given by \cref{mr} for the parameters used in our simulations, both for monomers ($\rho_\mathrm{l}=0.759\sigma^{-3}$, $h_t = 21.02\sigma$, $\nu = 2.41\sigma^2/\tau$, \cref{fig:1}a) and oligomers ($\rho_\mathrm{l}=1.06\sigma^{-3}$, $h_t = 20.22\sigma$, $\nu = 23.4\sigma^2/\tau$, \cref{fig:1}b). The time is shown in units of the characteristic time of the thin film equation, which is given by $t_{0,\mathrm{mono}}$ for monomers and by $t_{0,\mathrm{oligo}}$ for oligomers. As can be seen, $M_\mathrm{R}$ decays, on the time scale on which the film height evolves, very rapidly. Consequently, the Markovian approximation is valid, which explains the validity of the thin film equation. Interestingly, the decay is, relative to the characteristic time scale, significantly faster in the case of oligomers (\cref{fig:1}b) than in the case of monomers (\cref{fig:1}a). The most likely explanation for this is the fact that the viscosity, which by \cref{diffusionequation} determines both the time scale of the relaxation and the macroscopic characteristic time to be compared to\footnote{The exponential functions in \cref{mr} depend on the product $\nu s$. If we insert $s=\tilde{s}t_0$ with the dimensionless time $\tilde{s}$ and the macroscopic characteristic time $t_0$, which itself is proportional to $\nu$, this gives a factor $\nu^2 \tilde{s}$.}, is much higher in the case of polymers. Figure \eqref{fig:1} also suggests the possibility of approximating $M_\mathrm{R}$ by simpler functions (such as exponentials) in order to simplify numerical simulations based on \cref{governingequationmemory}.

\subsubsection{Validity of small gradient assumption}
An important approximation is the one made in \cref{twoapproximations}, namely that gradients of the diffusion tensor -- which, as can be seen from \cref{twodcurrent,diffusiontensor1}, is determined by the correlation of the $xy$-components of the particle velocities --  are smaller in the $x/y$-direction than in the $z$-direction. In \cref{fig:2}, we show numerical results for these velocity correlations obtained for both monomers and oligomers as a function of the distance in the $xy$ direction and in the $z$ direction. (Due to the rotational symmetry in the $xy$-plane, the $x$- and $y$-direction are equivalent.) As can be seen, the decay of the velocity correlations is much steeper for the $z$-direction than in the $x/y$-direction. Steeper decays imply larger gradients, which confirms the assumption made in the derivation.

\subsubsection{Test of prediction for memory function}
A central result of the analytical approach proposed here that goes beyond the established thin film equation is the prediction \eqref{mzt} for the form of the correlation function $M(z,t,s)$. To compare $M(z,t,s)$ obtained from the MD simulations with the approximated form \cref{mzt}, we first calculate $\hat{\vec{J}}_\mathrm{T}(z,t)$ via \cref{eq_J_T}. For this purpose, the MD system is binned in steps of $\SI{1}{\sigma}$ in $z$-direction starting from a height of $\SI{0}{\sigma}$. Afterwards, $C(z,z',t,s)$ is calculated via the definition \eqref{definitionofc} by taking the ensemble average over 1000 independent trajectories. 

\begin{figure}
    \centering
    \includegraphics[width=\linewidth]{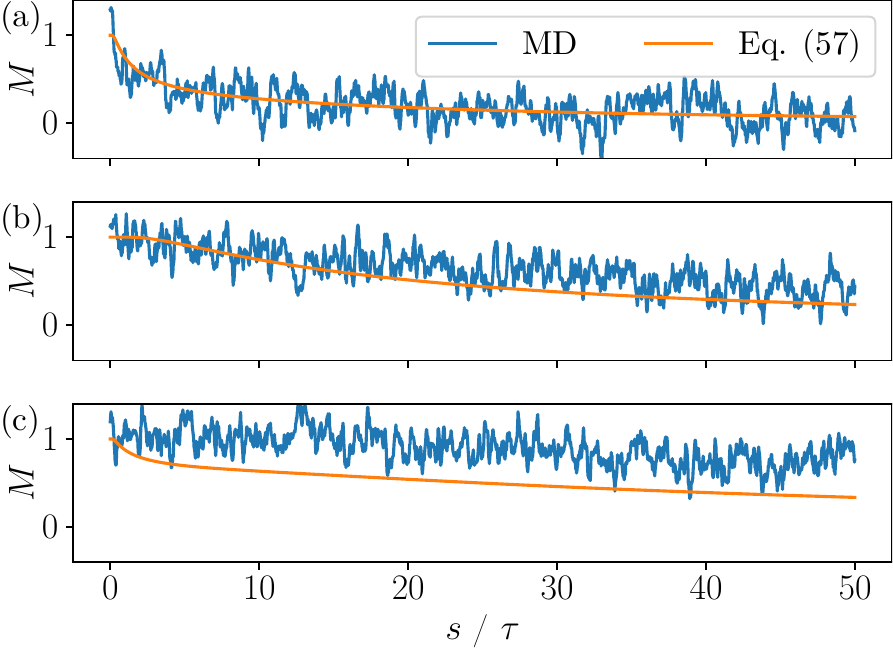}
    \caption{\label{fig:3}$M(z,t,s)$ calculated from MD simulations for a simple fluid via \cref{approximated2} compared to the approximated form \eqref{mzt} for slices at positions (a) $z= \SI{2.5}{\sigma}$, (b) $z= \SI{8.5}{\sigma}$, and (c) $z= \SI{18.5}{\sigma}$. The height $h$ is $\SI{21.0}{\sigma}$.}
\end{figure}

Figure \eqref{fig:3} shows the results for $M(z,t,s)$ as a function of $s$ computed from theory (orange) and simulation (blue) for the monomer system for three different values of $z$. The function $M$ decays, both in the theory and in the simulation, more rapidly for small values of $z$. As can be seen, the MD data, despite being very noisy, agree quite well with the approximated form \eqref{mzt}. A possible reason for deviations, especially for higher values of $z$, may be the absence of a sharp transition between liquid and vapor phases (which the theory assumes) in the simulations.

\begin{figure}
    \centering
    \includegraphics[width=\linewidth]{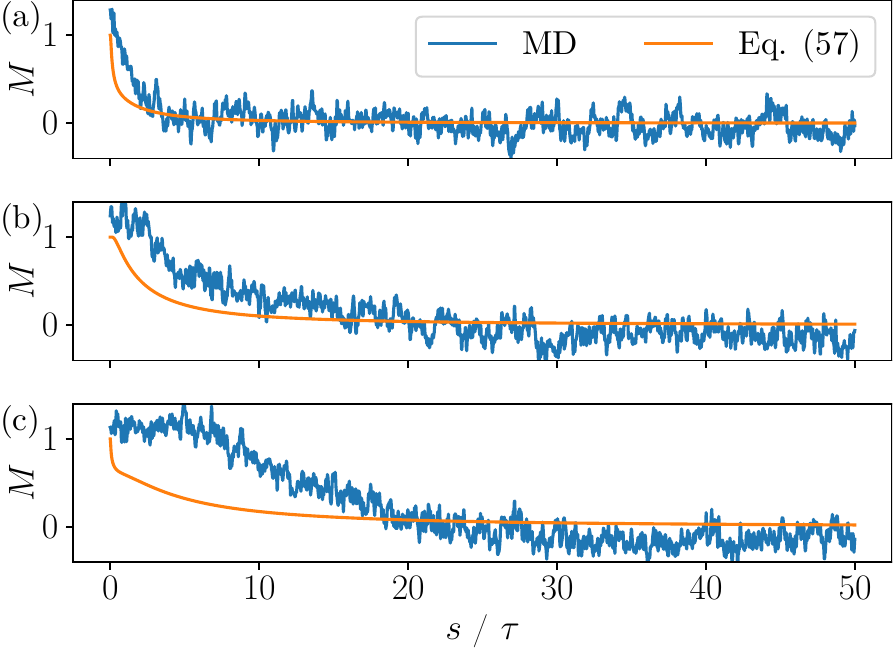}
    \caption{\label{fig:4}$M(z,t,s)$ calculated from MD simulations for a polymer via \cref{approximated2} compared to the approximated form \eqref{mzt} for slices at positions (a) $z= \SI{2.5}{\sigma}$, (b) $z= \SI{8.5}{\sigma}$, and (c) $z= \SI{18.5}{\sigma}$. The height $h$ is $\SI{20.2}{\sigma}$.}
\end{figure}

Figure \eqref{fig:4} shows the same plots for the oligomer system. Also here, the agreement between theory and simulation is relatively good, in particular for small values of $z$. However, it is not as good as in the case of monomers. A possible explanation for this is the fact that polymers have a complex viscoelastic dynamics (in contrast to the viscous dynamics of simple fluids) and that therefore the assumption that the correlation function is governed by \cref{diffusionequation} might be less accurate in the case of polymers. In particular, on short time scales the microscopic degrees of freedom in the polymer system might be governed by an effective viscosity that is smaller than the full viscosity of the system, as a consequence of which $M$ relaxes slower than predicted by our theoretical model. The reason for the dependence of the viscosity on the considered time scale is that the fact that the particles are connected in a chain has little influence on their dynamics on short time scales, but strongly affects the observed correlations on longer ones.


\section{\label{conclusion}Conclusions}
We have presented a microscopic derivation of the thin-film equation using the Mori-Zwanzig projection operator formalism. This method allows to obtain this result very directly, without the need to employ the full hydrodynamic equations as in standard derivations and leading to microscopic expressions for the transport coefficients and the memory functions. These are found to be in good agreement with simulation results. 

Our results contribute to the microscopic understanding of thin film hydrodynamics \cite{StienekerTGH2021,ToppSGH2022}. Moreover, they are of general interest for researchers working on projection operator methods, given that strategies for determining memory functions \cite{JungHS2017,JungHS2018,AyazSDN2022,MeyerPS2020,MeyerWSS2021} and the validity of the Markovian assumption \cite{teVrugt2022} have been widely studied recently. Possible further applications of our approach include the derivation of extensions of the standard thin-film equation incorporating, e.g., additional order parameter fields.

\acknowledgments{We thank Jens Bickmann, Svetlana V. Gurevich, Tobias Frohoff-H\"ulsmann, Moritz Stieneker, and Uwe Thiele for helpful discussions. M.t.V.\ and R.W.\ are funded by the Deutsche Forschungsgemeinschaft (DFG, German Research Foundation) -- Project-IDs 525063330 and 433682494 -- SFB 1459. M.t.V.\ also thanks the Studienstiftung des deutschen Volkes for financial support during the first part of the project.}

\nocite{apsrev41Control}
\bibliographystyle{apsrev4-1}
\bibliography{refs,control}
\end{document}